\documentstyle[prl,aps,preprint]{revtex}

\newcommand{\om}{\omega}
\newcommand{\tom}{\tilde \omega}
\newcommand{\tR}{\tilde R}
\newcommand{\pt}{\partial}
\tightenlines

\begin{document}

\draft              

\title{Vortex precession in a rotating nonaxisymmetric trapped
           Bose-Einstein condensate}

\author{Alexander L. Fetter$^{1,2,3}$ and Jong-kwan Kim$^{1,3}$}
\address{$^{1}$Geballe Laboratory for Advanced Materials,
                   Stanford University,
                   Stanford, CA 94305-4045\\
             $^{2}$Department of Physics,
                   Stanford University,
                   Stanford, CA 94305-4060\\
             $^{3}$Department of Applied Physics,
                   Stanford University,
                   Stanford, CA 94305-4090 }
\date{October 4, 2001}
\maketitle

\begin{abstract}
We study the precession of an off-axis straight vortex in a rotating
nonaxisymmetric harmonic trap in the Thomas-Fermi (TF) regime.
A time-dependent variational Lagrangian analysis yields
the dynamical equations of the vortex and the precessional angular velocity
in two-dimensional (2D) and three-dimensional (3D) condensates.
\end{abstract}

\pacs{PACS numbers: 03.75.Fi, 05.30.Jp}

\section{introduction}
\label{sec:intro}
Since the original experimental observation of Bose-Einstein
condensation in dilute atomic gases~\cite{Ande95,Brad95,Davi95},
the creation and detection of vortices has attracted great interest
(for a recent review article, see, for example Ref. \cite{Fett01}).
In particular, one experiment has measured the precession frequency
$\omega_p$
of a vortex line in a nonrotating stationary spherical condensate
\cite{Ande00} and compared $\omega_p$ to various theoretical estimates
\cite{Jack00,Lund00,Svid00a,Svid00b,McGe00,Fede01}.
Although these theoretical approaches largely predict the same results,
perhaps the most transparent and physical picture \cite{Lund00,Svid00a}
relies on a variational Lagrangian method \cite{Pere96}
that focuses directly on the position of the vortex core.
For an axisymmetric condensate in rotational equilibrium
at a small angular velocity $\Omega \ll \omega_{\perp}$,
the original precession frequency $\omega_p$ is altered to
$\omega_p(\Omega) = \omega_p-\Omega$ by the external rotation.
The present work describes the nontrivial generalization of the variational
Lagrangian method to a large rotating {\em nonaxisymmetric} condensate in the
Thomas-Fermi (TF) limit.  For extreme asymmetry, we may note that the
precession frequency
$\omega_p(\Omega)$ becomes nearly independent of $\Omega$ (for $\Omega \ll
\omega_{\perp}$) because the irrotational flow induced by the rotating
asymmetry screens the effect of the external rotation.

\section{precessional dynamics of a straight vortex}
\label{sec:precess}
A Bose condensate is described by a
macroscopic order parameter (the condensate wave function) $\Psi$, with
particle density
$n=|\Psi|^2$.  Consider a condensate in a nonaxisymmetric harmonic trap
potential
$V_{\rm tr}({\bf r}) = \frac{1}{2} M (\om_x^2 x^2+\om_y^2
y^2+\om_z^2 z^2)$.    The condensate experiences both the
trap potential
$V_{\rm tr}$ and the
self-consistent Hartree interaction
$V_H = g|\Psi|^2 = gn$ arising from the interaction with all the other
particles, where $g = 4\pi\hbar^2 a/M$ is the effective interparticle
interaction strength and $a > 0$ is the
$s$-wave scattering length~\cite{Dalf99,Fett99}. At zero temperature, the
order parameter obeys the time-dependent Gross-Pitaevskii (GP) equation

\begin{equation}
i\hbar\frac{\partial \Psi}{\partial t} = -\frac{\hbar^2 \nabla^2}{2M}\Psi +
V_{\rm tr}\Psi + g|\Psi|^2\Psi.\label{GP}
\end{equation}
When the trap  rotates with angular
velocity
$\Omega$ about the
$z$ axis, the corresponding  GP equation in a co-rotating frame acquires an
additional term $-\Omega L_z\Psi = i\hbar\, \Omega\, {\bf
r\times{\bbox\nabla}}\Psi$ on the right-hand side, arising from the
transformation to the rotating frame~\cite{Lifs}.

Instead of working directly with the time-dependent GP equation (\ref{GP}),
it is often preferable to introduce a Lagrangian formalism based on
the Lagrangian functional

\begin{equation}
\label{lall}
{\mathcal L}[\Psi] = {\mathcal T}[\Psi] - {\mathcal F}[\Psi],
\end{equation}
where
\begin{equation}
\label{ltime}
{\mathcal T}[\Psi] = \int dV\, \frac{i \hbar}{2}
                         \left( \Psi^* \frac{\pt \Psi}{\pt t}
                               - \Psi \frac{\pt \Psi^*}{\pt t}
                         \right)
\end{equation}
is the time-dependent part of the Lagrangian that is analogous to the
kinetic energy in classical particle mechanics.  Similarly,
   the
free energy ${\mathcal F}[\Psi] $ here plays the
role of the potential energy. In a nonrotating system, $\cal F$ is simply
the energy functional
$\cal E$, which includes the gradient (bending) energy, the trap energy,
and the self-consistent Hartree energy.  When the system rotates,
however, the free energy has an extra term
${\mathcal F} = {\mathcal E} - \Omega L_z$~\cite{Lifs},
and the appropriate free-energy  functional of the system is

\begin{equation}
\label{lenergy}
{\mathcal F}[\Psi] = \int dV \left(\frac{\hbar^2}{2M} |\nabla\Psi|^2
                                + V_{\rm tr} |\Psi|^2 + \frac{g}{2} |\Psi|^4
                                + i \hbar \Omega\, \Psi^*  \frac{\pt
\Psi}{\pt \phi}
                                 \right).
\end{equation}
Given the Lagrangian, the associated action is the time integral ${\cal
S}[\Psi] = \int_{t_1}^{t_2}  dt {\cal L}[\Psi]$, and it is straightforward
to verify that the action is stationary with respect to small variations
of
$\Psi$ and
$\Psi^*$ when $\Psi$ obeys the time-dependent GP equation (\ref{GP}). Thus
the time-dependent GP equation is the Euler-Lagrange equation for this
problem.

This Lagrangian formalism is exact, but it also provides the basis for a
powerful approximate variational method.  If the wave function $\Psi$
depends on various parameters, the integral in Eq.~(\ref{ltime}) will
involve the first time derivatives of these parameters.  Consequently, the
resulting  $\cal L$ will  serve as the effective Lagrangian
that provides the corresponding dynamical equations for these
parameters.  This method has been used to study the quadrupole
oscillations of a vortex-free condensate~\cite{Pere96}.  Here, we use it
to study the dynamics of a vortex in a rotating nonaxisymmetric trap.

Specifically, we assume that a straight singly quantized vortex is
displaced from the center of the trap with transverse coordinates $x_0$
and
$y_0$; these variables will serve as time-dependent variational
parameters that obey Lagrange's equations \cite{Pere96,Lund00}. In the TF
limit, the vortex induces negligible change in the condensate density,
and, in the presence of the rotation, $|\Psi|^2$ is given
by~\cite{Rica01,Madi01}
\begin{equation}
\label{tf1}
g|\Psi({\bf r})|^2 = \tilde{\mu} - \tilde{V}_{\rm tr}({\bf r})
              = \tilde{\mu} - \case{1}{2} M \left( \tom_x^2 x^2
                                                  + \tom_y^2 y^2
                                                  + \om_z^2 z^2\right)
\end{equation}
where $\tilde\omega_x$ and $\tilde\omega_y$ are effective oscillator
frequencies that determine the shape of the condensate in the rotating
asymmetric trap.  Specifically, they are given in terms of a parameter
$\alpha$:

\begin{equation}
\tom_x^2 = \om_x^2 + \alpha^2 - 2\alpha\Omega, \qquad
\tom_y^2 = \om_y^2 + \alpha^2 + 2\alpha\Omega,
\end{equation}
where   $\alpha$ satisfies a cubic equation

\begin{equation}
2\alpha^3 + \alpha(\om_x^2+\om_y^2-4\Omega^2) + \Omega(\om_x^2-\om_y^2) = 0.
\end{equation}
The parameter $\alpha$ has a simple physical
interpretation~\cite{Rica01}, for it
determines the  irrotational flow ${\bf v}_0$ induced by the rotating
nonaxisymmetric trap through  the  velocity  potential $\Phi_0 =
\alpha\,xy$,  with
${\bf v}_0 = \bbox{\nabla} \Phi_0$.  As noted below, $\alpha =
-\Omega\left(\tom_x^2-\tom_y^2\right)/\left(\tom_x^2+\tom_y^2\right)$.

     Equation (\ref{tf1}) shows that the condensate density has  the
familiar TF parabolic  form

\begin{equation}
\label{tf}
|\Psi|^2 = n_0 \left( 1 - \frac{x^2}{\tR_x^2}
                            - \frac{y^2}{\tR_y^2}
                            - \frac{z^2}{\tR_z^2} \right),
\end{equation}
if the right-hand side is positive and zero otherwise.  Here
$n_0 = \tilde{\mu} / g = M \tilde{\mu} / 4 \pi a \hbar^2$ is the
central density,
\ $\tR_i = \sqrt{2 \tilde{\mu} / M \tom_i^2}$ (for $i=x,y$),
$\tR_z=\sqrt{2\tilde{\mu}/M\om_z^2}$ are the TF radii, and
$\tilde\mu =
\case{1}{2}\hbar\tilde\omega_0\left(15 Na/\tilde d_0\right)^{2/5}$ with
$\tilde\omega_0^3 = \tom_x\tom_y\om_z$ and $\tilde
d_0^2=\hbar/M\tilde\omega_0$.
For a small angular velocity $\Omega\ll\om_\perp$
where $\om_\perp^2\equiv\case{1}{2}(\om_x^2+\om_y^2)$,
the renormalization of the  oscillator
frequencies  and TF  radii is of order $\Omega^2$ and thus
small.

In our variational approach, we assume that the dominant vortex-induced
contribution to the Lagrangian arises  from the condensate's superfluid motion
and  use the following trial wave function
\begin{equation}
\label{trial}
\Psi = |\Psi| e^{i (S_0 +  S_1)},
\end{equation}
where $|\Psi|$ follows from the TF density profile in
Eq.~(\ref{tf}).  The first term of the phase is taken as  $S_0=(M\alpha/\hbar)
\, xy$, which represents  the irrotational motion induced by the
rotating nonaxisymmetric trap (note that $\alpha$ vanishes if
$\omega_x=\omega_y$).  The second term is taken to be
\begin{equation}
\label{circ}
S_1 = \arctan \left( \frac{y-y_0}{x-x_0} \right),
\end{equation}
which characterizes the circulating flow around the vortex line.  In contrast
to the analysis of Refs.~\cite{Fett98,Lund00}, we do not include an image
vortex because the form  of the TF condensate density ensures that the
particle current vanishes automatically at the TF surface.

  Variation of the
functional (\ref{lenergy}) yields an Euler-Lagrange equation for
$S_0$, whose solution gives \cite{Svid00a,Rica01,Madi01,Fett99a,Fett74}
\begin{equation}
\label{phase}
S_0 = - \frac{M \Omega }{\hbar}
            \left(\frac{\tom_x^2-\tom_y^2}{\tom_x^2+\tom_y^2}\right) \,xy
        =   \frac{M \Omega }{\hbar}
            \left(\frac{\tR_x^2-\tR_y^2}{\tR_x^2+\tR_y^2}\right) \,xy.
\end{equation}
Note that the time-dependent parameters $x_0$ and $y_0$ appear only
in the phase $S_1$, so that
\begin{eqnarray}
\pt_t \Psi &=& i \, \Psi \pt_t S_1  \nonumber \\
               &=& -i\, \Psi \, \dot{\mathbf r}_0 \! \cdot \!
                   {\bbox\nabla} \,S_1 \nonumber  \\
               &=& -i\, \Psi \,
                   \frac{-\dot{x}_0(y-y_0) + \dot{y}_0(x-x_0)}
                        {(x-x_0)^2 + (y-y_0)^2} \, .
\end{eqnarray}

\section{Two-dimensional condensate}
\label{sec:2d}
For simplicity, we first consider a two-dimensional condensate that is
unbounded in the $z$ direction (taking $\omega_z\to 0$), in which case  the
various terms in the Lagrangian are interpreted per unit length. To evaluate
the time-dependent part (\ref{ltime}), it is convenient to introduce
dimensionless coordinates scaled with the renormalized TF radii and shift the
origin of coordinates to the position of the vortex. In this way, we find
\begin{eqnarray}
\label{2dtime}
{\cal T}[\Psi] &=&  \hbar n_0 \tR_x^2 \tR_y^2 \int d^2r\,
                 \frac{-\dot{x}_0 y + \dot{y}_0 x}{\tR_x^2 x^2 + \tR_y^2 y^2}
                 \left[1 - (r \cos \phi + x_0)^2-(r \sin \phi + y_0)^2 \right]
                 \nonumber \\
            &=&  \hbar n_0 \tR_x^2 \tR_y^2 \int d\phi\,
                 \frac{-\dot{x}_0 \sin \phi +\dot{y}_0 \cos \phi}
                      {\tR_x^2 \cos^2 \phi + \tR_y^2 \sin^2 \phi}
                 \left( - \frac{2}{3}B^3 - AB \right),
\end{eqnarray}
where
\begin{eqnarray}
A &\equiv& 1 - x_0^2 - y_0^2,  \nonumber \\
B &\equiv& x_0 \cos \phi + y_0 \sin \phi,
\end{eqnarray}
with $x_0$ and $y_0$   also dimensionless.  As anticipated in the general
discussion, $\cal T$ indeed depends on the first time derivatives of
the parameters $x_0$ and $y_0$.

Lagrange's equations now take the familiar form

\begin{equation}\label{lagr}
{\mathcal{D}}_x\left({\cal T}-{\cal F}\right) = 0, \quad
{\mathcal{D}}_y\left({\cal T}-{\cal F}\right) = 0,
\end{equation}
where

\begin{equation}
{\mathcal{D}}_x \equiv \frac{d}{dt} \frac{\pt}{\pt \dot{x}_0}
                           - \frac{\pt}{\pt x_0},
\end{equation}
and similarly for ${\mathcal{D}}_y$.
If this operator is applied to functions of the
form appearing in Eq.~(\ref{2dtime}), it is straightforward to verify that
\begin{equation}
{\mathcal{D}}_x \left[ (-\dot{x}_0 \sin \phi + \dot{y}_0 \cos \phi)
                           f(x_0,y_0,\phi)
                    \right]
          = - \dot{y}_0 \left( \cos \phi \, \pt_{x_0}
                             + \sin \phi \, \pt_{y_0} \right) f(x_0,y_0,\phi).
\end{equation}
In our case,
\begin{equation}
\left( \cos \phi \, \pt_{x_0} + \sin \phi \, \pt_{y_0} \right) A = -2B,
\qquad \left( \cos \phi \, \pt_{x_0} + \sin \phi \, \pt_{y_0} \right) B = 1.
\end{equation}
Applying this operator to the time-dependent term (\ref{2dtime})
simplifies the integral considerably  and yields
\begin{eqnarray}
{\mathcal{D}}_x {\cal T}[\Psi]
          &=& \dot{y}_0\, \hbar n_0 \tR_x^2 \tR_y^2
              \int d\phi\, \frac{A}{\tR_x^2 \cos^2 \phi + \tR_y^2 \sin^2 \phi}
              \nonumber \\
          &=& \dot{y}_0\, 2\pi \hbar n_0 \tR_x \tR_y \, (1-x_0^2-y_0^2).
\end{eqnarray}
The first of Lagrange's equations (\ref{lagr})  then becomes
\begin{equation}
{\mathcal{D}}_x {\mathcal T}[\Psi]
          = - \frac{\pt \triangle {\mathcal F}[\Psi]}{\pt x_0},
\end{equation}
where $\triangle {\mathcal F}$ denotes the extra free energy associated with
the presence of a vortex. This equation, along with a similar one for the $y_0$
dependence can be rewritten as
\begin{equation}
\label{Magnus}
2\pi \hbar \, n_0 \,\tR_x \tR_y \, \hat z \times \dot {\mathbf r}_0
       = {\bbox \nabla}_{0} \, \triangle {\mathcal F}({\mathbf r}_0),
\end{equation}
where ${\mathbf r}_0 = (x_0,y_0)$ is the dimensionless coordinate
vector of the displaced vortex. As is familiar in the context of
two-dimensional vortices in superfluid helium \cite{Hess67}, this
dynamical equation can be interpreted as a balance between the
Magnus force and the gradient of the free energy with respect to
the coordinates of the vortex \cite{Jack00,Lund00,McGe00}.

We can calculate the free-energy part with logarithmic accuracy~\cite{Fett98}
by substituting (\ref{trial})-(\ref{phase}) into the energy
functional (\ref{lenergy}):
\begin{equation}
\label{2denergy}
\triangle {\mathcal F}({\mathbf r}_0,\Omega)
       = 2\pi \tilde{\mu} \xi^2 n_0 (1 - r_0^2)
         \left[ \ln \left( \frac{\tR_\perp}{ \xi}\right)
               - \frac{M \Omega}{\hbar} \,
                 \frac{\tR_x^2 \tR_y^2}{\tR_x^2 + \tR_y^2} \,
                 (1 - r_0^2)
         \right] \, ,
\end{equation}
where $\tR_\perp^2 = 2 \tR_x^2 \tR_y^2/(\tR_x^2 + \tR_y^2)$
and $\xi^2 = \hbar^2 / 2M \tilde{\mu}$. With these results,
we can derive the equations of motion and obtain the
angular frequency $ \om_p(\Omega)$ of the vortex precession.  An
easy calculation yields
\begin{equation}
\label{2dresult}
\dot{x}_0 = -\om_p(\Omega) y_0,
\quad
\dot{y}_0 = + \om_p(\Omega) x_0,
\end{equation}
where
\begin{equation}
\label{2dfreq}
\om_p(\Omega) \equiv \om_p^0
                         - \frac{2 \tR_x \tR_y}{\tR_x^2 + \tR_y^2} \Omega
\end{equation}
and
\begin{equation}
\om_p^0 \equiv \frac{\hbar }{M \tR_x \tR_y}
                    \ln \left( \frac{\tR_\perp}{ \xi} \right) \frac{1}{1-r_0^2}
\end{equation}
is the precession frequency in a nonrotating condensate of stretched dimensions
$\tR_x$ and $\tR_y$.
For a not-too-large angular velocity $\Omega\ll\om_\perp$,
note that $\om_p^0\simeq\om_p(0)$.
An earlier more intricate derivation of this result
analyzed  the dynamical motion of each element of the vortex core
\cite{Svid00b}.  If $\om_p(\Omega)$ is positive, the vortex precesses in
the  positive (counterclockwise) sense, namely with the same sense
as the circulating fluid around the vortex.
Such behavior is indeed seen in the recent JILA experiments \cite{Ande00}
for nonrotating condensates.

For an axisymmetric condensate with $\tR_x=\tR_y=\tR_\perp$,
the precession frequency $\omega_p(0)$
in the absence of rotation is shifted to
$\om_p(\Omega) = \om_p^0 -\Omega$ by the applied rotation
(as expected from the transformation to the rotating frame \cite{Linn00}).
For an asymmetric trap, however, the induced irrotational flow
in Eq.~(\ref{phase}) acts to screen the effect of the external rotation.
In the extreme limit $\tR_y \ll \tR_x$, for example, we have
$\om_p(\Omega) \approx \om_p^0 - (2\tR_y / \tR_x)\Omega$,
so that
the precession then becomes nearly independent of $\Omega$ for small $\Omega$.

This behavior can be understood
by examining the background fluid flow velocity in the rotating frame
${\mathbf v} = (\hbar/M){\bbox\nabla}'S_0 - \Omega\hat z\times {\mathbf r}'$
in the absence of the vortex~\cite{Fett74}; for clarity, we now use
primes to denote the original dimensional coordinates with,
for example, $x'= x \tR_x$.
In particular, $v_x(x',y') = 2\Omega \,y' \, \tR_x^2/(\tR_x^2+\tR_y^2)$.
This induced velocity appears in the dimensional equation of motion as
\begin{equation}
\frac{\dot x_0'}{\tR_x}
      = - \om_p^0 \frac{y_0'}{\tR_y}
        + \frac{v_x(x_0',y_0')}{\tR_x}
      = - \om_p^0 \frac{y_0'}{\tR_y}
        + \frac{2\Omega \tR_x}{\tR_x^2+\tR_y^2} \, y_0',
\end{equation}
which readily reproduces the first of Eqs. (\ref{2dresult}),
and similarly for the second equation.

\section{Three-dimensional condensate}
\label{sec:3d}
It is not difficult to generalize these results to a three-dimensional
TF condensate. For example, the explicitly time-dependent part of
the Lagrangian now contains an additional integral over the axial coordinate
$z$:
\begin{eqnarray}
{\mathcal T}[\Psi] &=& \hbar n_0 \tR_x^2 \tR_y^2 \tR_z
                \int d^3r\, \frac{- \dot{x}_0 y + \dot{y}_0 x}
                                 {\tR_x^2 x^2 + \tR_y^2 y^2}
                \left[ 1 - (r \cos \phi + x_0)^2
                         - (r \sin \phi + y_0)^2 -z^2 \right] \nonumber \\
            &=& 2 \hbar n_0 \tR_x^2 \tR_y^2 \tR_z
                  \int d^2r \int_0^{z_{\rm max}} dz\,
                  \frac{-\dot{x}_0y + \dot{y}_0x}{\tR_x^2 x^2 + \tR_y^2 y^2}
                  \left[ 1 - (r \cos \phi + x_0)^2
                           - (r \sin \phi + y_0)^2 -z^2 \right]  \nonumber \\
            &=& \frac{ \hbar n_0 \tR_x^2 \tR_y^2 \tR_z }{6}
                \int d\phi\, \frac{-\dot{x}_0 \sin \phi + \dot{y}_0 \cos \phi}
                                {\tR_x^2 \cos^2 \phi + \tR_y^2 \sin^2 \phi}
                \nonumber  \\
            & & \phantom{\frac{q \hbar n_0 R_x^2 R_y^2 R_z }{6} \int d\phi}
                \times\left\{-5 B A^{3/2} - 3 B^3 A^{1/2}
                       + 3\left(A + B^2 \right)^2
                          \left[\frac{\pi}{2}
                                + \arctan \left( \frac{B}{-A^{1/2}} \right)
                          \right]
                \right\},
\end{eqnarray}
where $z_{\rm max}^2 = 1 - (r \cos \phi + x_0)^2 - (r \sin \phi + y_0)^2$.
Although this angular integral is difficult to evaluate directly,
only the simpler quantity ${\mathcal{D}}_x {\mathcal T}[\Psi]$ is needed,
and a straightforward analysis yields
\begin{eqnarray}
{\mathcal{D}}_x {\mathcal T}[\Psi]
          = \dot{y}_0 \frac{8 \pi \hbar n_0}{3} \tR_x \tR_y \tR_z
(1-r_0^2)^{3/2}.
\end{eqnarray}
In addition, the free energy is given by \cite{Svid00a}
\begin{equation}
\triangle {\mathcal F}({\mathbf r}_0,\Omega)
	= \frac{8 \pi \tilde{\mu} \xi^2 n_0 \tR_z}{3}
          (1-r_0^2)^{3/2}
          \left[ \ln \left( \frac{\tR_\perp}{ \xi}\right)
                 - \frac{4}{5} \frac{M \Omega}{\hbar}
                               \frac{\tR_x^2 \tR_y^2}{\tR_x^2 + \tR_y^2}
                               (1 - r_0^2)
	  \right].
\end{equation}
The dynamical equation again reduces to (\ref{Magnus}), and the only
difference is that the angular frequency $\omega_p^0$ has a modified
numerical coefficient
\begin{equation}
\label{3dfreq}
{\omega}_p(\Omega)
	\equiv \frac{3}{2} \frac{\hbar}{M \tR_x \tR_y}
	         \ln \left( \frac{\tR_\perp}{ \xi} \right) \frac{1}{1-r_0^2}
               - \frac{2 \tR_x \tR_y}{\tR_x^2 + \tR_y^2} \Omega;
\end{equation}
this expression  agrees with  an earlier result based on the method of
matched asymptotic expansions \cite{Svid00a}.

Note that our analysis for a three-dimensional condensate
considers only a straight vortex line.  In the present Thomas-Fermi limit
of a large condensate, the axis of the vortex line must lie
perpendicular to   the condensate's surface~\cite{Fett01,Svid00b}.
Hence the approximation of a straight vortex is directly applicable to the
central region of a disk-shaped condensate.   For cigar-shaped
condensates, in contrast, the bending of the vortex generally plays an
essential role.  Indeed, for sufficiently elongated condensates, a
straight vortex on the central axis can become unstable with respect to
bending deformations~\cite{Svid00b,Fede01}.

\section{Discussion and conclusions}
The Lagrangian functional (\ref{lall}) contains two parts,
${\mathcal T}[\Psi]$ with explicit time dependence and ${\mathcal F}[\Psi]$
involving the time-independent free energy in the rotating frame.
If only the latter quantity is considered (which is the Hamiltonian in
the rotating frame), the stability can be inferred directly by
considering how $\mathcal F$ changes for small lateral displacements of the
vortex from the central position. Such methods have been used to
analyze the onset of metastability in a rotating nonaxisymmetric TF
condensate \cite{Fett01,Svid00a}.

In contrast, the full Lagrangian allows a more complete dynamical description.
In the present case, the time dependence of the vortex position
${\mathbf r}_0$ yields equations of motion (\ref{Magnus}), which readily give
the precession frequency $\omega_p(\Omega)$ in a rotating condensate.
As mentioned in Sec.~{\ref{sec:precess}}, 
the present description does not include an image
vortex. Although such an image vortex  is needed to satisfy the appropriate
boundary condition for a condensate with uniform density in a rigid
container~\cite{Fett99a}, the present TF condensate density vanishes at the TF
boundary so that the superfluid current automatically vanishes there.
In addition, the image vortex has negligible effect on the energy
when calculated with logarithmic accuracy.
As emphasized in Refs. \cite{Fett01,McGe00},
the vortex is unstable (metastable) whenever the gradient of
$\Delta {\mathcal F}({\mathbf r}_0,\Omega)$  with
respect to ${\mathbf r}_0$  in Eq.~(\ref{Magnus}) acts to move the vortex
away from (back toward) the center of the trap.

The present Lagrangian approach gives a direct  physical
derivation of the precession frequency. Although the results
describe only a single straight vortex displaced from the $z$ axis,
the analysis is far
simpler than the method of matched asymptotic expansions applied
to the full Gross-Pitaevskii equation~\cite{Svid00a,Svid00b}. For
a one-component condensate, it fully justifies the intuitive
dynamical equation (\ref{Magnus}) suggested earlier by McGee and
Holland~\cite{McGe00}. It would be valuable to generalize the present
approach to the more interesting and  challenging case of two components,
where additional restoring (buoyancy) forces have been proposed~\cite{McGe00}.

\acknowledgments
This work has benefited from many helpful discussions with
M.~J.~Holland and A.~A.~Svidzinsky. It has been supported
in part by the NSF, Grant No.~99-71518.

\end{document}